\documentclass[twocolumn, reprint, letterpaper,aps,prl,superscriptaddress,nofootinbib,floatfix, showpacs]{revtex4-2}

\usepackage{graphicx}% Include figure files
\usepackage{dcolumn}% Align table columns on decimal point
\usepackage{bm}% bold math
\usepackage{hyperref}
\usepackage[T1]{fontenc}
\def\MJD{\textsc{Majorana Demonstrator}}
\def\dem{\textsc{Demonstrator}}
\def\MJ{\textsc{Majorana}}
\def\nnbb{$0\nu\beta\beta$}
\def\tnbb{$2\nu\beta\beta$}
\def\Qbb{$Q_{\beta\beta}$}
\def\mbb{$m_{\beta\beta}$}

\begin{document}

\preprint{APS/123-QED}

\title{Final Result of the \bfseries{\scshape{Majorana Demonstrator}}'s Search for Neutrinoless Double-$\beta$ Decay in $^{76}$Ge}

\newcommand{\ITEP}{National Research Center ``Kurchatov Institute'' Institute for Theoretical and Experimental Physics, Moscow, 117218 Russia}
\newcommand{\JINR}{Joint Institute for Nuclear Research, Dubna, 141980 Russia} 
\newcommand{\lbnl}{Nuclear Science Division, Lawrence Berkeley National Laboratory, Berkeley, CA 94720, USA}
\newcommand{\lbnle}{Engineering Division, Lawrence Berkeley National Laboratory, Berkeley, CA 94720, USA}
\newcommand{\lanl}{Los Alamos National Laboratory, Los Alamos, NM 87545, USA}
\newcommand{\queens}{Department of Physics, Engineering Physics and Astronomy, Queen's University, Kingston, ON K7L 3N6, Canada}
\newcommand{\uw}{Center for Experimental Nuclear Physics and Astrophysics, and Department of Physics, University of Washington, Seattle, WA 98195, USA}
\newcommand{\unc}{Department of Physics and Astronomy, University of North Carolina, Chapel Hill, NC 27514, USA}
\newcommand{\duke}{Department of Physics, Duke University, Durham, NC 27708, USA}
\newcommand{\ncsu}{Department of Physics, North Carolina State University, Raleigh, NC 27695, USA}	
\newcommand{\ornl}{Oak Ridge National Laboratory, Oak Ridge, TN 37830, USA}
\newcommand{\ou}{Research Center for Nuclear Physics, Osaka University, Ibaraki, Osaka 567-0047, Japan}
\newcommand{\pnnl}{Pacific Northwest National Laboratory, Richland, WA 99354, USA}
\newcommand{\ttu}{Tennessee Tech University, Cookeville, TN 38505, USA}
\newcommand{\sdsmt}{South Dakota Mines, Rapid City, SD 57701, USA}
\newcommand{\usc}{Department of Physics and Astronomy, University of South Carolina, Columbia, SC 29208, USA}
\newcommand{\usd}{Department of Physics, University of South Dakota, Vermillion, SD 57069, USA}  
\newcommand{\ut}{Department of Physics and Astronomy, University of Tennessee, Knoxville, TN 37916, USA}
\newcommand{\tunl}{Triangle Universities Nuclear Laboratory, Durham, NC 27708, USA}
\newcommand{\mpi}{Max-Planck-Institut f\"{u}r Physik, M\"{u}nchen, 80805, Germany}
\newcommand{\tum}{Physik Department and Excellence Cluster Universe, Technische Universit\"{a}t, M\"{u}nchen, 85748 Germany}
\newcommand{\williams}{Physics Department, Williams College, Williamstown, MA 01267, USA}
\newcommand{\ciemat}{Centro de Investigaciones Energ\'{e}ticas, Medioambientales y Tecnol\'{o}gicas, CIEMAT 28040, Madrid, Spain}
\newcommand{\iu}{Department of Physics, Indiana University, Bloomington, IN 47405, USA}
\newcommand{\iuceem}{IU Center for Exploration of Energy and Matter, Bloomington, IN 47408, USA}

\author{I.J.~Arnquist}\affiliation{\pnnl} 
\author{F.T.~Avignone~III}\affiliation{\usc}\affiliation{\ornl}
\author{A.S.~Barabash}\affiliation{\ITEP}
\author{C.J.~Barton}\affiliation{\usd}
\author{P.J.~Barton}\affiliation{\lbnl}
\author{K.H.~Bhimani}\affiliation{\unc}\affiliation{\tunl} 
\author{E.~Blalock}\affiliation{\ncsu}\affiliation{\tunl} 
\author{B.~Bos}\affiliation{\unc}\affiliation{\tunl} 
\author{M.~Busch}\affiliation{\duke}\affiliation{\tunl}	
\author{M.~Buuck}\altaffiliation{Present address: SLAC National Accelerator Laboratory, Menlo Park, CA 94025, USA}\affiliation{\uw} 
\author{T.S.~Caldwell}\affiliation{\unc}\affiliation{\tunl}	
\author{Y-D.~Chan}\affiliation{\lbnl}
\author{C.D.~Christofferson}\affiliation{\sdsmt} 
\author{P.-H.~Chu}\affiliation{\lanl} 
\author{M.L.~Clark}\affiliation{\unc}\affiliation{\tunl} 
\author{C.~Cuesta}\affiliation{\ciemat}	
\author{J.A.~Detwiler}\affiliation{\uw}	
\author{Yu.~Efremenko}\affiliation{\ut}\affiliation{\ornl}
\author{H.~Ejiri}\affiliation{\ou}
\author{S.R.~Elliott}\affiliation{\lanl}
\author{G.K.~Giovanetti}\affiliation{\williams}  
\author{M.P.~Green}\affiliation{\ncsu}\affiliation{\tunl}\affiliation{\ornl}   
\author{J.~Gruszko}\affiliation{\unc}\affiliation{\tunl} 
\author{I.S.~Guinn}\affiliation{\unc}\affiliation{\tunl} 
\author{V.E.~Guiseppe}\affiliation{\ornl}	
\author{C.R.~Haufe}\affiliation{\unc}\affiliation{\tunl}	
\author{R.~Henning}\affiliation{\unc}\affiliation{\tunl}
\author{D.~Hervas~Aguilar}\affiliation{\unc}\affiliation{\tunl} 
\author{E.W.~Hoppe}\affiliation{\pnnl}
\author{A.~Hostiuc}\affiliation{\uw} 
\author{M.F.~Kidd}\affiliation{\ttu}	
\author{I.~Kim}\affiliation{\lanl} 
\author{R.T.~Kouzes}\affiliation{\pnnl}
\author{T.E.~Lannen~V}\affiliation{\usc} 
\author{A.~Li}\affiliation{\unc}\affiliation{\tunl} 
\author{A.M.~Lopez}\affiliation{\ut}	
\author{J.M. L\'opez-Casta\~no}\affiliation{\ornl} 
\author{E.L.~Martin}\altaffiliation{Present address: Duke University, Durham, NC 27708}\affiliation{\unc}\affiliation{\tunl}	
\author{R.D.~Martin}\affiliation{\queens}	
\author{R.~Massarczyk}\affiliation{\lanl}		
\author{S.J.~Meijer}\affiliation{\lanl}	
\author{S.~Mertens}\affiliation{\mpi}\affiliation{\tum}		
\author{T.K.~Oli}\affiliation{\usd}  
\author{G.~Othman}\altaffiliation{Present address: Universit{\"a}t Hamburg, Institut f{\"u}r Experimentalphysik, Hamburg, Germany}\affiliation{\unc}\affiliation{\tunl} 
\author{L.S.~Paudel}\affiliation{\usd} 
\author{W.~Pettus}\affiliation{\iu}\affiliation{\iuceem}	
\author{A.W.P.~Poon}\affiliation{\lbnl}
\author{D.C.~Radford}\affiliation{\ornl}
\author{A.L.~Reine}\affiliation{\unc}\affiliation{\tunl}	
\author{K.~Rielage}\affiliation{\lanl}
\author{N.W.~Ruof}\affiliation{\uw}	
\author{D.C.~Schaper}\affiliation{\lanl} 
\author{D.~Tedeschi}\affiliation{\usc}		
\author{R.L.~Varner}\affiliation{\ornl}  
\author{S.~Vasilyev}\affiliation{\JINR}	
\author{J.F.~Wilkerson}\affiliation{\unc}\affiliation{\tunl}\affiliation{\ornl}    
\author{C.~Wiseman}\affiliation{\uw}		
\author{W.~Xu}\affiliation{\usd} 
\author{C.-H.~Yu}\affiliation{\ornl}
\author{B.X.~Zhu}\altaffiliation{Present address: Jet Propulsion Laboratory, California Institute of Technology, Pasadena, CA 91109, USA}\affiliation{\lanl} 

\collaboration{{\sc{Majorana}} Collaboration}
\noaffiliation


\newcommand{\ackfunding}{This material is based upon work supported by the U.S.~Department of Energy, Office of Science, Office of Nuclear Physics under contract / award numbers DE-AC02-05CH11231, DE-AC05-00OR22725, DE-AC05-76RL0130, DE-FG02-97ER41020, DE-FG02-97ER41033, DE-FG02-97ER41041, DE-SC0012612, DE-SC0014445, DE-SC0018060, DE-SC0022339, and LANLEM77/LANLEM78. We acknowledge support from the Particle Astrophysics Program and Nuclear Physics Program of the National Science Foundation through grant numbers MRI-0923142, PHY-1003399, PHY-1102292, PHY-1206314, PHY-1614611, PHY-1812409, PHY-1812356, PHY-2111140, and PHY-2209530. We gratefully acknowledge the support of the Laboratory Directed Research \& Development (LDRD) program at Lawrence Berkeley National Laboratory for this work. We gratefully acknowledge the support of the U.S.~Department of Energy through the Los Alamos National Laboratory LDRD Program and through the Pacific Northwest National Laboratory LDRD Program for this work.  We gratefully acknowledge the support of the South Dakota Board of Regents Competitive Research Grant. We acknowledge the support of the Natural Sciences and Engineering Research Council of Canada, funding reference number SAPIN-2017-00023, and from the Canada Foundation for Innovation John R.~Evans Leaders Fund.  This research used resources provided by the Oak Ridge Leadership Computing Facility at Oak Ridge National Laboratory and by the National Energy Research Scientific Computing Center at Lawrence Berkeley National Laboratory, a U.S.~Department of Energy Office of Science User Facility. We thank our hosts and colleagues at the Sanford Underground Research Facility for their support.}

\date{\today}

\begin{abstract}
  The \MJD\ searched for neutrinoless double-$\beta$ decay (\nnbb) of $^{76}$Ge using modular arrays of high-purity Ge detectors operated in vacuum cryostats in a low-background shield.
  The arrays operated with up to 40.4~kg of detectors (27.2~kg enriched to $\sim$88\% in $^{76}$Ge).
  From these measurements, the \dem\ has accumulated $64.5$~kg~yr of enriched active exposure.
  With a world-leading energy resolution of 2.52~keV FWHM at the 2039~keV \Qbb\ (0.12\%), we set a half-life limit of \nnbb\ in $^{76}$Ge at $T_{1/2}>8.3\times10^{25}$~yr (90\% C.L.).
  This provides a range of upper limits on \mbb\ of $(113-269)$~meV (90\% C.L.), depending on the choice of nuclear matrix elements. 
\end{abstract}

\maketitle

Neutrinoless double-beta decay (\nnbb) is a hypothetical nuclear process involving the unbalanced creation of two new matter particles but no antimatter~\cite{Agostini2022, Dolinski2019, Barabash2018, Vergados2016, DelloOro2016}.
This lepton-number-violating process is predicted generically by many grand-unification theories, as well as by leptogenesis~\cite{Fukugita1986}, a leading explanation of why the universe is matter-dominated.
A half-life measurement is sensitive to the Majorana mass of the electron neutrino, \mbb, and requires nuclear matrix elements that must be calculated through many body nuclear theory~\cite{Engel2017, Ejiri2019}.
The experimental signature of \nnbb\ is a peak in total electron kinetic energy at the Q-value (\Qbb) of the decay.
While double-beta decay with the emission of two neutrinos (\tnbb) has been directly observed in 9~isotopes~\cite{Barabash2020}, no experiment has yet seen \nnbb. % I only count 9 in Barabash's paper
A robust program involving many low-background experiments has searched for \nnbb\ in multiple isotopes, with half-life limits surpassing $10^{25-26}$~yr in some cases~\cite{Majorana2019, Gerda2020, KamLANDZen2022, Exo2019, CUORE:2021mvw}.
For example, the GERDA experiment has established the leading half-life limit in $^{76}$Ge of $1.8\times10^{26}$~yr~\cite{Gerda2020}.

The \MJD\ searched for \nnbb\ in $^{76}$Ge~\cite{Majorana2014} using high purity germanium (HPGe) detectors enriched in this isotope.
Since its most recent published limit using data collected from 2015--2018~\cite{Majorana2019}, the experiment has more than doubled its previous enriched active detector exposure from 26.1~kg~yr to 64.5~kg~yr.
HPGe detectors are a mature technology offering excellent energy resolution and low intrinsic background~\cite{Avignone2019}.
Germanium can be enriched to a high isotopic fraction of $^{76}$Ge, which acts as both source and detector, enabling a high detection efficiency.
The \MJD\ consisted of two modules, each with an array of HPGe detectors operated in a vacuum cryostat.
The cryostats and structural components were fabricated from ultra-low background copper that was electroformed underground (UGEFCu)~\cite{Majorana:ugefcu} and carefully selected plastics.
Low background front end electronics, cables, and connectors were developed for the experiment~\cite{Majorana:electronics, Majorana:HV}.
The modules were enclosed by a multi-layered low-background shield, consisting of, from inner to outer layers, 5~cm of UGEFCu, 5~cm of commercially sourced copper, and 45~cm of high-purity lead.
This shield construct was enclosed in an aluminum radon exclusion box that was constantly purged with low-Rn liquid nitrogen (LN) boil-off gas.
The Rn-exclusion box was surrounded by an active muon veto consisting of 32 plastic scintillating panels with nearly $4\pi$ coverage~\cite{Majorana:muonveto}.
The vacuum and cryogenic hardware and control electronics were located just outside this enclosure and connected to the cryostats via a shielded crossarm penetrating the other layers.
Finally, the entire assembly was enclosed in 5~cm borated and 25~cm pure polyethylene neutron shielding.
A $^{228}$Th line source for each module was stored outside the shield, with a penetration and helical track enabling its deployment for detector calibrations~\cite{Majorana:calibration}.
The experiment was located in the Davis campus on the 4850-foot level (4300~m.w.e.) of the Sanford Underground Research Facility (SURF) in Lead, South Dakota~\cite{Heise:2015}.
To ensure low backgrounds, components of the modules and shield were subjected to an extensive radioassay campaign~\cite{Majorana:assay}.
To reduce cosmogenic activation, time spent on the surface for detectors and detector components was minimized and logged in a database~\cite{Majorana:ptdb}.

Three novel HPGe detector geometries were used, each with a p-type, point-like electrode centered on one face and an n-type electrode covering the remaining surfaces, separated by a thin passivated surface.
The \dem\ primarily used p-type, point contact (PPC) detectors~\cite{Luke1989, Barbeau2007} enriched to $87.4\pm0.5\%$ in $^{76}$Ge and having masses of $0.6-1.2$~kg.
We also used Broad Energy Germanium (BEGe\textsuperscript{TM}) Detectors~\cite{Canberra:bege} made of natural Ge with typical masses of $\sim$0.6~kg,
and inverted coaxial point contact (ICPC) detectors~\cite{Cooper2011} enriched to $88\pm1\%$ in $^{76}$Ge, with masses of $1.3-2.1$~kg.
Module~1 began operation in July~2015 and Module~2 began in Aug.~2016.
The modules were installed with 35 PPC detectors (29.7~kg) and 23 BEGe detectors (14.4~kg), of which up to 22.1~kg of enriched detectors and 10.0~kg of natural detectors were operational.
The most common reason for non-operating detectors was failure in the high voltage and signal electronics cable connectors.
In Nov.~2019, Module~2 was removed from the shield and upgraded with improved cables, connectors and shielding; as a result, all Module~2 detectors were operational during subsequent run time.
Five PPC detectors (5.5~kg) were removed for early testing in the LEGEND-200 experiment~\cite{Legend2017} and replaced with four ICPC detectors (6.7~kg).
After this, up to 27.2~kg of enriched detectors and 13.2~kg of natural detectors were operational.

Detector signals were digitized using ADCs developed for the GRETINA Experiment~\cite{Vetter2000}, with 14~bit resolution and a sampling rate of 100~MS/s~\cite{Anderson2009}.
Each detector was read out with two gains, a high gain channel with dynamic range up to $\sim$3~MeV, and a low gain channel extending up to $\sim$10~MeV.
The high gain channels were used for most of the analysis; use of low gain channels is reserved for events with no high gain waveform available, due to dead time after a noise trigger or rejection of the high gain channel, and for high energy events~\cite{Majorana:trinuc, Majorana:alpha-n}.
An internal trapezoidal filter was used to trigger each ADC channel independently, recording samples in a $20.2-38.2~\mu$s acquisition window to disk.
A pulser signal was fed to the front-end electronics every 8~s to monitor the electronics stability and detector livetime.

%We collected data nearly continuously from Jan.~2016 to Mar.~2021. % Jason: can comment back in if there is space (redundant with earlier text)
During blinded operation, 75\% of physics runs were blinded by restricting access to data files, with cycles of 31~h of open data followed by 93~h of blind data.
Typical trigger rates during physics runs for the full array were around 70~Hz, dominated by near-threshold and pulser events.
Physics run data are divided into 13 datasets based on changes to the experimental configuration, described in the Supplemental Material \cite{Supplemental}.
Once per week, the $^{228}$Th line sources were deployed for 60--90~min, with trigger rates of up to 3000~Hz.
Every 2--3 months a long calibration run with $\gtrsim$18~h of data was collected for each module. 
In Jan.~2019, a $^{56}$Co line source was inserted into each track, and calibration data were collected for one week with each module.
This dataset has multiple double-escape peaks (DEPs) and single-escape peaks (SEPs) in an energy range bracketing \Qbb\, between 1500~keV and 2500~keV. 
These act as proxies for \nnbb\ events and were used for systematics measurements.

The digitized waveforms are used to calculate event energies and pulse-shape discrimination (PSD) parameters that are used to reject likely backgrounds.
%90\% of \nnbb\ events are localized within a few mm of a decay site located in the bulk of a detector, so background rejection techniques sought to remove events with differing topologies. % Jason: can comment back in if there is space (redundant with later text)
In this analysis, we use a set of improved analysis routines compared to the analyses described in Refs.~\cite{Majorana2018, Majorana2019}, and we will note these changes.
In addition, the PPC and ICPC detectors utilize subtly different digital signal processing algorithms; we describe the PPC parameters first, followed by the ICPC parameters.

\begin{table}[tb]
  \centering
  \begin{tabular}{lcc}
\hline\hline
 & PPCs & ICPCs \\
\hline
Total Exposure & $67.94$~kg yr & $3.12$~kg yr \\
Active Exposure & $61.64^{+0.89}_{-1.17}$~kg yr & $2.82^{+0.04}_{-0.05}$~kg yr \\
FWHM@2039~keV & $2.52\pm0.08$~keV & $2.55\pm0.09$~keV \\
$^{76}$Ge Enrichment & $87.4\pm0.5\%$ & $88.0\pm1.0\%$ \\
PSD Eff. & $86.1\pm3.9\%$ & $81.0^{+5.3}_{-7.3}\%$ \\
~~~Data Cleaning & $99.9\pm0.1\%$ & $99.9\pm0.1\%$ \\
~~~Low AvsE* Cut & $89.9^{+3.3}_{-3.2}\%$ & $85.2^{+4.2}_{-5.9}\%$ \\
~~~DCR Cut & $98.5\pm0.7\%$ & $97.9\pm1.1\%$ \\
~~~High AvsE* Cut & $97.9\pm1.0\%$ & $97.8\pm1.4\%$ \\
~~~Late Charge Cut & $99.3\pm0.7\%$ & $99.5^{+0.5}_{-0.9}\%$ \\
Containment Eff. & $90.8\pm1.3\%$ & $91.9\pm0.8\%$ \\
ROI Peak Eff. & $86.3\pm1.1\%$ & $86.9\pm1.2\%$ \\

\hline\hline
\footnotesize $^\ast$$A/E$ in the case of ICPCs \\
\end{tabular}

  \caption{\label{tab:efficiencies} A summary of key analysis parameters. The FWHM and efficiency values represent an exposure-weighted average across datasets. The PSD efficiencies are calculated for cuts applied in the order listed, and are applied multiplicatively. The ROI peak efficiency uses the optimal ROI~\cite{Agostini2017} defined for the Feldman-Cousins limit.}
\end{table}

%Energy reconstruction
Event energies are reconstructed from waveform amplitudes which are measured using a pole-zero (PZ) corrected trapezoidal filter.
Before applying any digital filters, waveforms are corrected for digitizer non-linearity~\cite{Majorana:adcnonlin}.
The energy is measured from the amplitude of the filtered waveform at a fixed time after the start time ($t_0$) of the signal.
By optimizing the PZ time constant for energy resolution, we minimize energy degradation due to charge trapping inside of the detectors~\cite{Majorana:chargetrapping}.
When analyzing energy calibrations, we model gamma peak shapes using the sum of a gaussian component and an exponentially-modified gaussian low-energy tail~\cite{Majorana2019}.
Peak shape parameters are determined using a simultaneous fit of the eight most prominent gamma-rays between 238--2615~keV.
Energy gains are calculated for each weekly calibration using the 2615~keV peak from the $^{228}$Th spectra and fixing the zero-ADC energy at 0~keV.
For each full dataset, a fine-tuned correction to the energy is obtained by combining all calibration runs and performing a new simultaneous peak-fit, with peak positions fit to a quadratic function of energy.
The exposure-weighted average FWHM at 2039~keV across all PPC detectors, including broadening due to gain drift and energy nonlinearities, is $2.52\pm0.08$~keV.
Since previous data releases~\cite{Majorana2018, Majorana2019}, the leading-edge fitter for finding $t_0$ was improved to use an asymmetric trapezoidal filter that preserves information in the rising edge, and $t_0$ was corrected for energy-dependent trigger walk; in addition, the quadratic energy correction is new to this analysis.
These changes correct nonlinearities in the energy response, especially at energies $<$100~keV, and reduce the low-energy tail.

\nnbb\ events predominantly occur in the bulk region of enriched detectors, with the charge cloud localized within $\sim$1~mm of the decay site.
Background rejection cuts are applied to remove events with multi-site topologies, events on the surfaces of detectors, and non-physical events.
Detector triggers within a rolling 4~$\mu$s window are grouped into events, and events with a detector multiplicity $>1$ are rejected.
The muon veto system triggers when a PMT signal above threshold is measured in two or more veto panels simultaneously, and the pulse amplitude of all panels are recorded.
A separate analysis identifies muon candidate events when at least two muon veto panels on different surfaces surpass an analysis threshold, while rejecting events triggered by LED pulsers used to monitor the panels.
Any events within 20~ms before and 1~s after a muon candidate are tagged and rejected~\cite{Majorana:muonveto, Majorana:muonactivity}.
All events for a module are rejected during periods of increased microphonic noise while LN dewars were refilled.
Multiple data cleaning PSD cuts are performed, accepting >99.9\% of signal-like events, to reject non-physical waveforms, pileup waveforms, and pulser events.

%AVSE CUT HERE
The point-contact detector geometries have relatively fast charge collection and a highly localized weighting potential, allowing for the identification and rejection of events that deposit energy in multiple sites within a detector, such as Compton-scattered gammas.
The parameter $AvsE$ represents a comparison between the maximum amplitude (A) of a waveform current pulse and the total energy (E).
Multi-site events have lower A for a given E than single-site events~\cite{Majorana:avse}.
$AvsE$ is tuned with long $^{228}$Th calibration runs to accept 90\% of events in the predominantly-single-site $^{208}$Tl DEP at 1592~keV.
Compared to previous data releases~\cite{Majorana2018, Majorana2019}, $AvsE$ has undergone several upgrades.
First, the $^{56}$Co calibration spectrum is used to measure and correct the width-energy dependence of the $AvsE$ distribution.
Second, we linearly correct for the correlation of $AvsE$ with the drift time, measured using the $0-90\%$ rise time ($\Delta t_{0-90}$).
The effect of these improvements is an improved signal acceptance and a more accurate determination of the single-site acceptance at \Qbb; based on $^{56}$Co studies performed after publication of Refs.~\cite{Majorana2018, Majorana2019}, the acceptance of the $AvsE$ cut in those analyses was $\sim$84\%.

%high AvsE cut
Surface events very close to the point contact produce faster-rising waveforms than bulk events~\cite{Majorana:tube}.
A cut value at high values of $AvsE$ is selected to accept $98\%$ of bulk events near \Qbb\ in the $^{228}$Th Compton-continuum from calibration data.
This background cut was not applied in Refs.~\cite{Majorana2018, Majorana2019}, and relies on the improvements made to the $AvsE$ parameter.

\begin{figure*}[tb]
  \centering
  \includegraphics[width=0.9\linewidth]{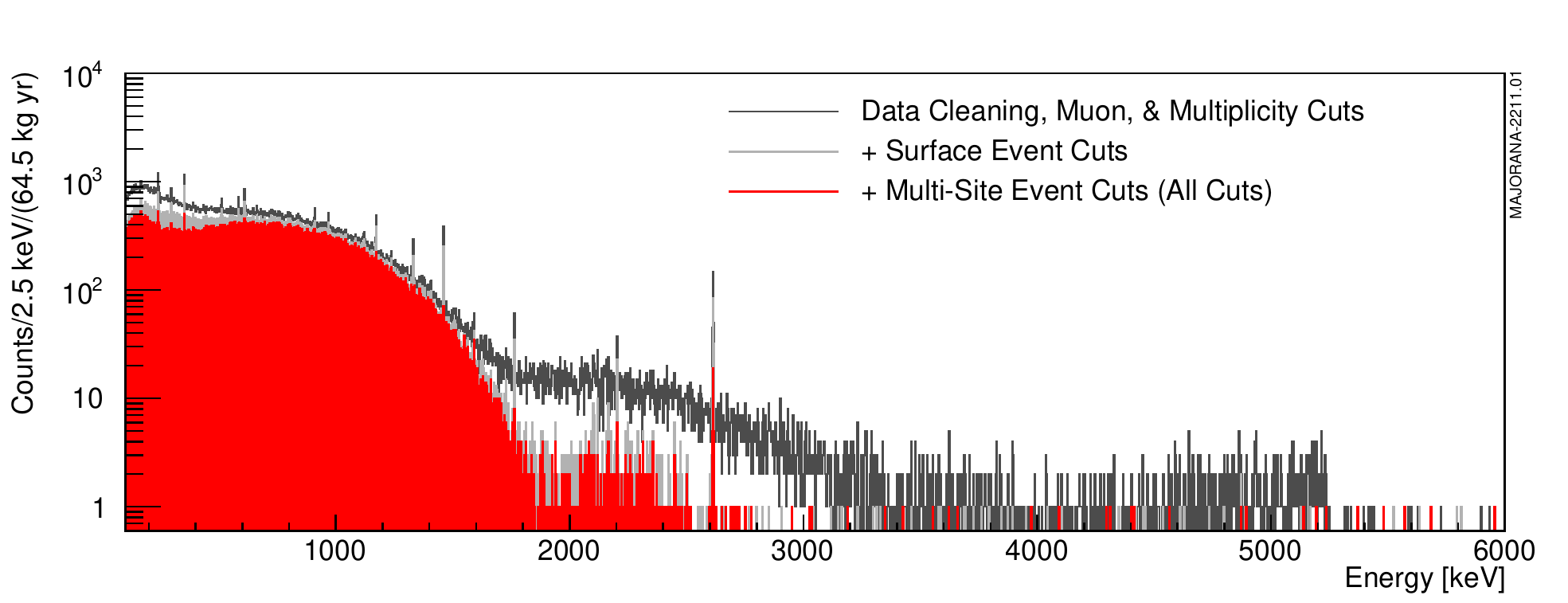}
  \caption{\label{fig:full_spectrum} The measured energy spectrum above 100~keV for the full enriched exposure after applying multiplicity and data cleaning cuts (dark gray), DCR, high-$AvsE$ and LQ cuts (light gray), and the low-$AvsE$ cut (red).}
\end{figure*}

%DCR cut
Alpha particles impinging on the passivated surface experience increased charge trapping and surface charge-collection effects that degrade their reconstructed energy, becoming an important source of background.
Much of this charge is collected slowly, increasing the slope of the falling tail of the waveforms relative to bulk events~\cite{Majorana:tube}.
Delayed charge recovery (DCR) is a measure of this slope increase, and can be used for surface alpha rejection.
The DCR cut is tuned using Compton-continuum events in $^{228}$Th calibration data to accept $99\%$ of bulk events, assuming it is normally distributed, in the energy range of $2028-2050$~keV bracketing \Qbb.
Since previous data releases~\cite{Majorana2018, Majorana2019}, the DCR parameter has been refined to improve the stability of the parameter with energy and time, and a correction for linear correlations with drift time has been added.

%LQ cut
Finally, events with a partial charge deposition in the transition layer between the n-type surfaces and the detector bulk experience energy degradation, and have waveforms with a slow-rise component.
We calculated the Late Charge (LQ) parameter, which is the integral of uncollected charge after a waveform has reached 80\% of its maximum value.
The slow-rise component of these waveforms increases LQ, and we cut events that fall $5\sigma$ above the center of the parameter distribution.
LQ is tuned using the 1592-keV DEP in $^{228}$Th calibration data, and is corrected for linear correlations with the drift time.
The parameter acceptance is measured to be $99.3\%$ using this same DEP, corrected for energy dependence using multiple DEPs in $^{56}$Co data.

Because the installed ICPC detectors are 2--3 times larger than PPC detectors, they experience greater charge trapping and diffusion effects.
To achieve similar energy and PSD performance to the PPC detectors, further improvements to the algorithms are implemented.
In place of $\Delta t_{0-90}$, we estimate drift time using the effective mean rise time, calculated as the integral of uncollected charge from $t_0$ until $1.5~\mu$s later.
Energy is calculated using a PZ correction and trapezoidal filter; these filters, however, are not optimized for charge trapping as in the PPC detectors.
Instead, we correct for charge trapping by correcting the energy for correlations with drift time that are either quadratic or linear depending on which achieved better resolution.
The ICPC detectors have an exposure-weighted average energy resolution of $2.55\pm0.09$~keV FWHM.
Instead of $AvsE$, we calculate the ratio between $A$ and $E$ ($A/E$)~\cite{Budjas2009}, and linear corrections for both drift time and energy are applied to this ratio.
We cut on both low and high values of $A/E$, for multi-site and near-point-contact events, respectively.
As a ratio, the width of $A/E$ increases rapidly at low energies and is only well understood for hits above 1~MeV, in contrast with $AvsE$; however, the improved drift time correction performs excellently near \Qbb, with only $4.76\pm0.50\%$ of events in the inherently-multi-site $^{208}$Tl SEP at 2103~keV surviving.
The DCR and LQ parameters are calculated similarly for ICPC and PPC detectors, applying corrections for linear correlations with effective mean rise time, rather than $\Delta t_{0-90}$.
The DCR parameter is additionally corrected for residual ADC nonlinearity which produces an energy dependent oscillation in the mean of the distribution.

The signal-efficiency for each PSD cut is measured using weekly calibrations, and averaged together for each full dataset.
Because of the differences between PPC and ICPC detectors, signal-efficiencies are calculated separately for each using identical procedures to measure these values, except where noted.
If a detector channel exhibits large instabilities between two calibrations, we reject its events for that period.
Uncertainty estimates account for statistics and systematic effects due to variations in cut parameters between weekly calibrations, energy dependencies of the cuts, and simulated differences between the waveform populations used to measure the signal-efficiency and \nnbb\ events.
The only difference applies to the low $A/E$ cut in ICPC detectors: no $^{56}$Co data is available for the ICPC detectors and thus no energy-width correction is applied to the parameter; instead, we reduce the signal acceptance from 90.2\% to 85.2\%, based on the estimated shift in signal-acceptance for the uncorrected parameter in PPC data.
The exposure-weighted averages of signal-efficiencies across all datasets are listed in Tab.~\ref{tab:efficiencies}.

The total exposure for PPC and ICPC detectors of 71.1~kg~yr is calculated by multiplying the full detector masses and the full livetime for each detector, after removing periods for which data from a detector are rejected.
We compute the active exposure by subtracting sources of dead time, including 0.77\% loss from periods of increased microphonic noise during LN dewar fills, 0.04\% loss from the period of each muon veto cut, and 0.51\% loss from retriggering dead time.
Detector volume with incomplete charge collection within $\sim$1~mm of detector surfaces is also subtracted from active exposure.
The fraction of active volume in PPC detectors is $92.0^{+1.3}_{-1.7}\%$, while the fraction in ICPC detectors is $90.9^{+1.2}_{-1.6}\%$.
The upper uncertainty estimates account for measurement uncertainty, and the lower uncertainty estimates also incorporate possible growth in dead layers while detectors are warm, which is being further investigated.
The total active exposure for PPC and ICPC detectors is $64.5^{+0.9}_{-1.2}$~kg~yr.
The active exposure from the datasets reported in Ref.~\cite{Majorana2019} increased from 26.0~kg~yr to 27.2~kg~yr in PPC detectors for the current analysis, thanks to added livetime due to an improved understanding of the stability of PSD parameters and the inclusion of data that were previously rejected by automated procedures for having high noise but manually verified to have a near-threshold rate comparable to accepted runs.

Unblinding of data proceeded in a staged fashion with basic data quality assurance checks performed at each stage.
No changes to analysis parameters or run selection were made after opening the energy window 1950--2350~keV which contains \Qbb.
Fig.~\ref{fig:full_spectrum} shows the energy spectrum, with the effect of cuts visible.
The spectrum is dominated by \tnbb\ below \Qbb, and near \Qbb\ most events are removed by the various cuts.
A background index is calculated using counts that pass all cuts within a 360-keV background estimation window, excluding $\pm5$~keV around the 2039~keV \Qbb-value and expected background gamma-rays at 2103~keV, 2118~keV and 2204~keV.
The background rate within this 360-keV window is assumed to be flat.
The surface cuts (DCR, high-$AvsE$ and LQ) remove 85\% of events in the background region.
The multi-site cut (low-$AvsE$) removes 49\% of the remaining events.
The majority of surface events are removed by more than one of the surface cuts; still, each uniquely cuts a significant population of events.
A parallel analysis using an interpretable boosted decision tree to leverage multi-variate correlations between these parameters achieves a similar result, with the potential for future background rejection~\cite{Majorana:bdt_analysis}.

\begin{figure}[tb]
  \centering
  \includegraphics[width=\linewidth]{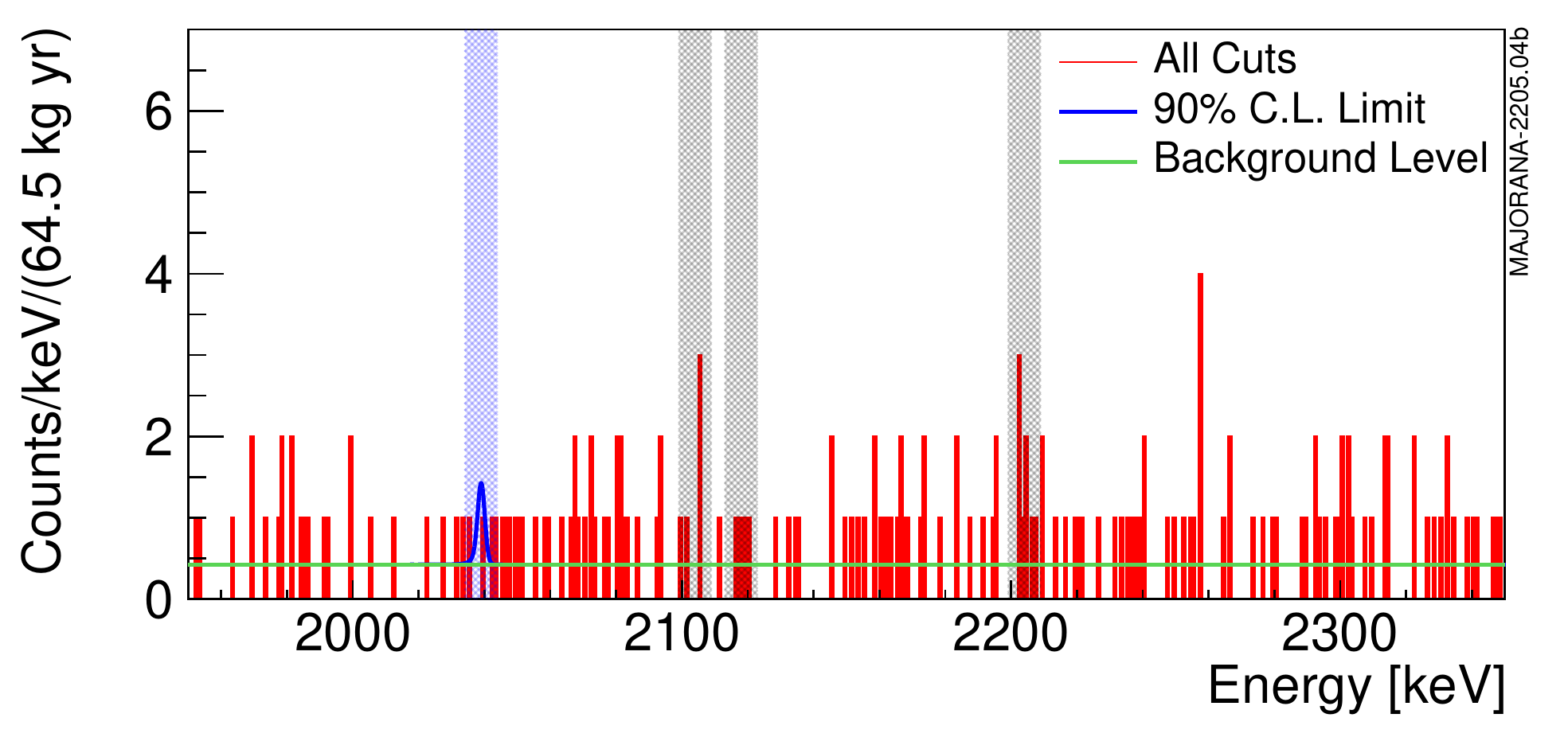}
  \caption{\label{fig:roi_spectrum} The measured energy spectrum for the full enriched exposure in the range of 1950-2350~keV, after applying all background cuts. The 360-keV background estimation window excludes the shaded 10~keV windows around expected gamma lines in gray, and the 10~keV window around the 2039~keV \Qbb\ in blue. Lines are drawn representing the background index (green) and a peak at the 90\% C.L. limit (blue).}
\end{figure}

After cuts, 153~events remain in the background estimation window, resulting in a background index of $16.6^{+1.4}_{-1.3}\times10^{-3}$~cts/(FWHM~kg~yr).
We also define a low background dataset, excluding the first dataset taken with Module~1 from June~2015 through Oct.~2015 prior to the installation of the inner Cu shield.
This low background dataset has an active exposure of 63.3~kg~yr in PPC and ICPC detectors, and 142-events in the background estimation window, resulting in a background index of $15.7^{+1.4}_{-1.3}\times10^{-3}$~cts/(FWHM~kg~yr).
We measure a significant difference in the background index between Module~1 ($18.1^{+1.8}_{-1.7}\times10^{-3}$~cts/(FWHM~kg~yr)) and Module~2 ($8.7^{+2.0}_{-1.7}\times10^{-3}$~cts/(FWHM~kg~yr)).
The \MJD\ measures a higher background index than predicted by the initial assay-based projections of ${<}2.5\times10^{-3}$~counts/(FWHM~kg~yr)~\cite{Majorana:assay}.
Current evidence suggests that the main source of this excess is located far from the HPGe detectors, most likely near the interface of the Module~1 crossarm and cryostat.
To better identify and measure this source, we are further analyzing possible background sources in a detailed model of the apparatus using the Geant4~\cite{Geant4} based package MaGe~\cite{Majorana:mage}.
Upon unblinding the 10~keV window centered on 2039~keV, 4 events were observed, consistent with the background expectation.
The events in the $1950-2350$~keV window are shown in Fig.~\ref{fig:roi_spectrum}.

A lower half-life limit for \nnbb\ is calculated using
\begin{equation}
  \label{eq:halflife}
  T_{1/2}>\mathrm{ln}(2)\frac{NT\epsilon_{tot}}{S}
\end{equation}
where $NT$ is the number of $^{76}$Ge nuclei in the active mass, $\epsilon_{tot}=0.78$ is the signal detection efficiency, and $S$ is the upper limit on the signal counts based on the observed data.
A limit of $T_{1/2}>8.3\times10^{25}$~yr is derived using an unbinned, extended profile likelihood method implemented in RooStats~\cite{Verkerke2003, Schott2011}; this is the same technique used for the limits in Refs.~\cite{Majorana2018,Majorana2019}.
Systematic uncertainties are included for the detection efficiency, exposure and peak shape, as listed in Tab.~\ref{tab:efficiencies}.
Because this limit is statistics dominated, these uncertainties have a very small effect on the result.
For this method, we use a 370-keV-wide fitting window, which includes the 360-keV background estimation window and the 10~keV window around 2039~keV.
A 90\% C.L. median sensitivity for exclusion of $T_{1/2}>8.1\times10^{25}$~yr is also derived using this construction.

Several additional statistical techniques were examined.
The Feldman-Cousins technique~\cite{Feldman1998} was implemented using a 3.8-keV-wide optimal region of interest (ROI) at \Qbb~\cite{Agostini2017}, with peak detection efficiency listed in Tab.~\ref{tab:efficiencies}.
This ROI contains 1~count, with an expectation of 1.52 background counts, producing a limit of $T_{1/2}>7.2\times10^{25}$~yr.
Two Bayesian analyses were implemented using Markov Chain Monte Carlo simulations performed with RooStats using the same likelihood function as for the frequentist result.
Using a flat prior on $1/T_{1/2}$ we calculate a half-life limit of $7.0\times10^{25}$~yr at 90\% C.I.
Using the Jeffreys prior, flat on $1/\sqrt{T_{1/2}}$, we calculate a limit of $1.1\times10^{26}$~yr at 90\% C.I.
A modified profile likelihood analysis using the more conservative CL$_S$ method~\cite{Read2000} to mitigate the effect of background down-fluctuations yields a limit of $6.4\times10^{25}$~yr at 90\% C.L.
The data required to perform a statistical analysis of the \MJD, including a full list of analysis parameters and events 370-keV fitting window, can be found in the Supplemental Materials~\cite{Supplemental}.

\begin{figure}
  \centering
  \includegraphics[width=\linewidth]{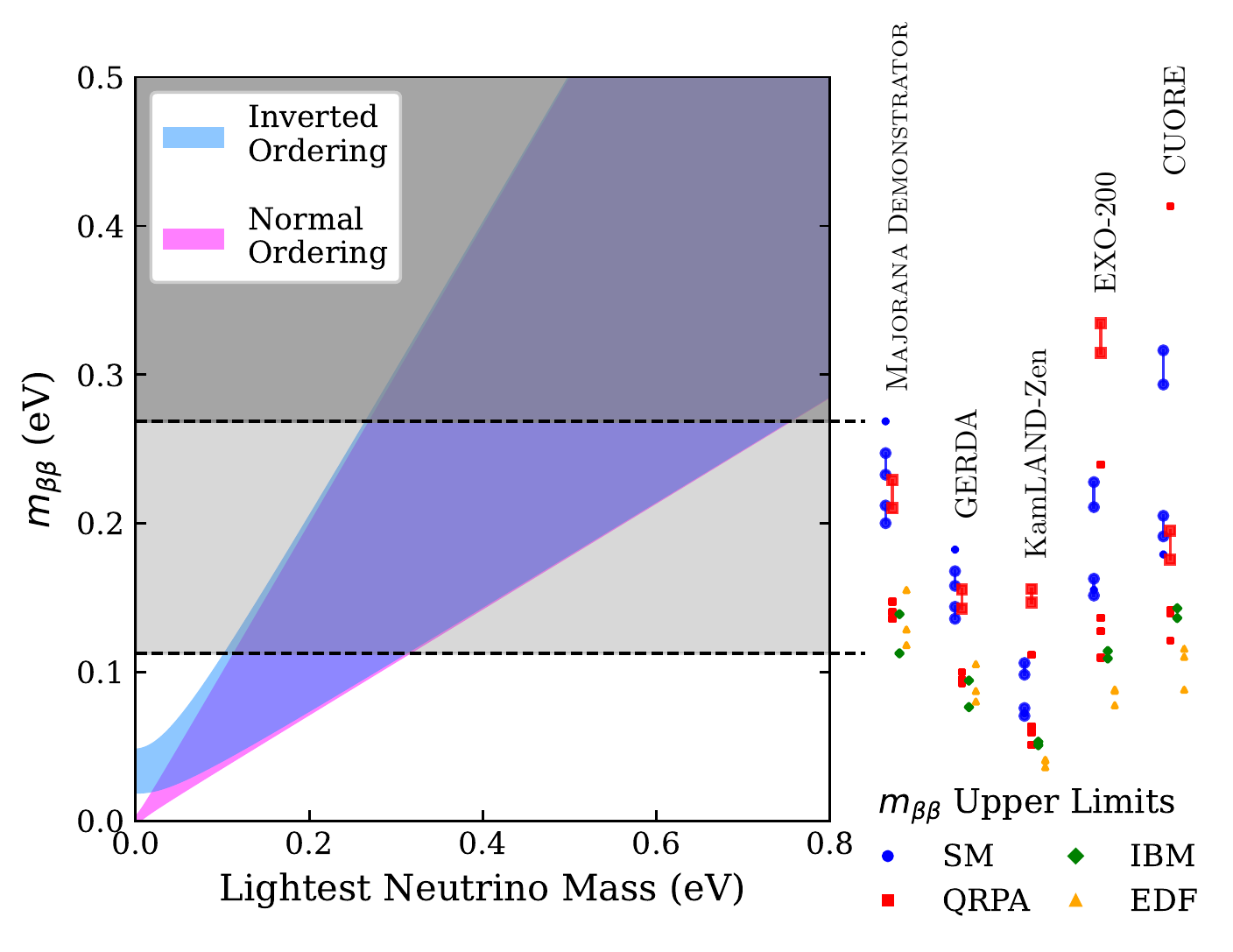}
  \caption{\label{fig:phase_space} Allowed values of \mbb\ for varying masses of the lightest neutrino eigenstate in the normal and inverted mass orderings, using the best-fit values of neutrino oscillation parameters~\cite{PDG2022}. Upper limits on \mbb\ at 90\% C.I. are calculated using half-life limits for the \MJD\ (this work), GERDA~\cite{Gerda2020}, KamLAND-Zen~\cite{KamLANDZen2022}, EXO-200~\cite{Exo2019}, and CUORE~\cite{CUORE:2021mvw}, using nuclear matrix elements calculated using the Quasi-Random Phase Approximation (QRPA)~\cite{Mustonen2013, Hyvarinen2015, Simkovic2018, Fang2018, Terasaki2020}, Shell Model (SM)~\cite{Menendez2017, Horoi2016, Coraggio2020}, Interacting Boson Model (IBM)~\cite{Barea2015, Deppisch2020}, and Energy Density Functional (EDF) theory~\cite{Rodriguez2010, Vaquero2013, Song2017}. The light gray shaded region shows the range of 90\% C.I. upper limits reported in this work, and the dark gray shows values of \mbb\ excluded for all nuclear matrix elements. }
\end{figure}

To place limits on \mbb\ assuming light neutrino exchange, we apply matrix element ($M_{0\nu}$) and phase space integral ($G_{0\nu}$) calculations.
We use a range of $M_{0\nu}$ values of $2.66-6.34$~\cite{Menendez2017, Horoi2016, Coraggio2020, Mustonen2013, Hyvarinen2015, Simkovic2018, Fang2018, Rodriguez2010, Vaquero2013, Song2017, Barea2015, Deppisch2020}, phase space factors ($G_{0\nu}$) of $2.36\times10^{-15}$/yr~\cite{Kotila2012} or $2.37\times10^{-15}$/yr~\cite{Mirea2015}, and an effective axial weak coupling of $g_A^{eff}=1.27$ for free nucleons~\cite{Ejiri2019}.
Applying these to the limit of $T_{1/2}>8.3\times10^{25}$, we calculate upper limits on \mbb\ in the range $(113-269)$~meV.
Fig.~\ref{fig:phase_space} shows this limit, along with allowed values of \mbb.
These limits are subject to theoretical uncertainties on $M_{0\nu}$ and $g_A^{eff}$~\cite{Ejiri2019}.

The \MJD\ has completed measurements with its enriched detectors.
Among \nnbb\ searches, the \dem\  achieved the best energy resolution of $\sim$2.5~keV at \Qbb, including for the ICPC detectors that will be predominantly used in LEGEND~\cite{Legend:pCDR}.
For this result, we have also demonstrated several new experimental techniques that will play an important role in future efforts, including the use of $^{56}$Co to correct energy dependent systematics in PSD cuts and the introduction of a novel PSD cut that is sensitive to events with charge deposition in n-type surfaces.
Through advances in ultra-low background materials, the \dem\ also achieved the second lowest background index normalized for energy resolution; only GERDA achieved lower backgrounds by immersing HPGe detectors in an active liquid argon shield~\cite{Gerda2020}.
The techniques used by both experiments to achieve low backgrounds are complementary and will be combined by LEGEND to achieve lower background rates than either experiment individually.
This has enabled \MJ\ to achieve limits comparable to other \nnbb\ experiments, even with a comparatively small exposure.
Since both the \MJD\ and GERDA achieved quasi-background free spectra near \Qbb, a combined analysis would measure a limit near the sum of the individual limits.

The authors appreciate the technical assistance of S.~Adair, J.F.~Amsbaugh, J.~Bell, T.H.~Burritt, G.~Capps, K.~Carney, J.~Cox, R.~Daniels, L.~DeBraeckeleer, C.~Dunagan, G.C.~Harper, C. Havener, G. Holman, R. Hughes, K.~Jeskie, K.~Lagergren, D.~Lee, O.~Loken, M.~Middlebook, A.~Montoya, A.W.~Myers, D.~Peterson, D.~Reid, L.~Rodriguez, H.~Salazar, A.R.~Smith, G.~Swift, J.~Thompson, P.~Thompson, M.~Turqueti, C.~Tysor, T.D.~Van Wechel, R.~Varland, T.~Williams, R.~Witharm, and H.~Yaver.
\ackfunding

\bibliographystyle{apsrev4-2}
\bibliography{mjd2022}

\end{document}